\documentclass[pdflatex,sn-mathphys-num]{sn-jnl}

\usepackage{bm}

\usepackage{graphicx}%
\usepackage{multirow}%
\usepackage{amsmath,amssymb,amsfonts}%
\usepackage{amsthm}%
\usepackage{mathrsfs}%
\usepackage[title]{appendix}%
\usepackage{xcolor}%
\usepackage{textcomp}%
\usepackage{manyfoot}%
\usepackage{booktabs}%
\usepackage{algorithm}%
\usepackage{algorithmicx}%
\usepackage{algpseudocode}%
\usepackage{listings}%

\theoremstyle{thmstyleone}%
\theoremstyle{thmstyletwo}%

\theoremstyle{thmstylethree}%

\begin{document}

\title[Article Title]{Enhancing Knowledge Tracing through Leakage-Free and Recency-Aware Embeddings}


\author*[1,2]{\fnm{Yahya} \sur{Badran}}\email{yahya.badran@h-ka.de}

\author*[1,2]{\fnm{Christine} \sur{Preisach}}\email{christine.preisach@h-ka.de}

\affil*[1]{\orgdiv{Institute Intelligent Systems Research Group}, \orgname{University of Applied Sciences Karlsruhe}, \orgaddress{\street{Moltkestr. 30}, \city{Karlsruhe}, \postcode{76133}, \state{Baden-Württemberg}, \country{Germany}}}

\affil[2]{\orgdiv{Faculty of Natural and Social Sciences}, \orgname{Pädagogische Hochschule Karlsruhe}, \orgaddress{\street{Bismarckstraße 10}, \city{Karlsruhe}, \postcode{76133}, \state{Baden-Württemberg}, \country{Germany}}}


\abstract{Knowledge Tracing (KT) aims to predict a student's future performance based on their sequence of interactions with learning content. Many KT models rely on knowledge concepts (KCs), which represent the skills required for each item. However, some of these models are vulnerable to label leakage, a phenomenon in which the input data inadvertently reveal the correct answer, particularly in datasets with multiple KCs per question.

We propose a straightforward yet effective solution to prevent label leakage by masking ground-truth labels during input embedding construction whenever such leakage could occur. To accomplish this, we introduce a dedicated \texttt{MASK} label, inspired by masked language modeling (e.g., BERT), to replace ground-truth labels. In addition, we introduce Recency Encoding, which encodes the step-wise distance between the current item and its most recent previous occurrence. This distance is important for modeling learning dynamics such as forgetting, which is a fundamental aspect of human learning, yet it is often overlooked in existing models. Recency Encoding demonstrates improved performance over traditional positional encodings on multiple KT benchmarks.

We show that incorporating our embeddings into KT models such as DKT, DKT+, AKT, and SAKT consistently improves prediction accuracy across multiple benchmarks. The approach is both efficient and widely applicable.

}

\keywords{Knowledge Tracing, Label Leakage, Deep Learning, Sequence Modeling, Knowledge Concepts}



\maketitle

\section{Introduction}\label{sec1}

Knowledge tracing (KT) models play a crucial role in personalization within intelligent tutoring systems (ITSs). These models estimate the evolving mastery state of a student regarding the various skills or concepts covered in coursework. Such skills are represented as Knowledge Concepts (KCs), and each item or question within an ITS typically involves multiple KCs required to respond correctly. For instance, a simple arithmetic problem like "$4 - 2 + 3 = ?$" could involve two KCs: "summation" and "subtraction". \footnote{The presented paper is an extended version of \cite{csedu25}}

Given the large number of questions in an ITS and the limited number of interactions per student, many KT models leverage KCs to mitigate the inherent sparsity of student-item interactions \cite{pykt,akt}. This is typically achieved by expanding each question into its constituent KCs, so that a single interaction involving a multi-KC question is replaced by multiple interactions, one for each KC. This transformation mitigates the effects of data sparsity by exploiting KCs shared across multiple questions.

Despite these advantages, using KC sequences introduces the issue of label leakage \cite{pykt}. This occurs when models inadvertently exploit shared labels among KCs originating from the same question to infer the ground-truth label. For example, consider a question associated with two KCs, $KC_1$ and $KC_2$. Because both KCs originate from the same student response, they receive the same correctness label. After expansion, the question is represented as two separate KC-level interactions. When predicting the label for $KC_2$, the model may exploit the observed label associated with $KC_1$ from the same question, rather than relying on the student's preceding interaction history.

Since ground-truth labels are unavailable during deployment, label leakage can artificially inflate evaluation metrics \cite{pykt}. In contrast, classical models such as Bayesian Knowledge Tracing (BKT) \cite{bkt} avoid this issue by modeling each KC independently. However, their performance generally falls short compared to deep learning-based KT models which are capable of capturing complex dependencies among KCs \cite{Howdeep,whendeep,DKTforget}.

In this paper, we address the label leakage problem and further enhance the performance of deep learning-based knowledge tracing models solely through modifications to their input embeddings, without altering the underlying model architecture. Specifically, we introduce two complementary embedding strategies: \emph{leakage-free embeddings}, which prevent potentially leaked labels, and \emph{recency-aware embeddings}, which incorporate KC-specific recency information.

To prevent label leakage, we introduce a special \texttt{MASK} label that replaces ground-truth labels in cases where leakage is likely, particularly when multiple KCs belong to the same question. This mask label is inspired by techniques from masked language modeling \cite{BERT}. The \texttt{MASK} label is integrated directly into the input embeddings and applied during both training and inference.


Beyond addressing label leakage, input embeddings are further enhanced with a novel recency encoding method to capture temporal learning patterns. The expanded KC sequence is leveraged to encode recency information for each KC individually. Specifically, the encoding represents the number of time steps since each KC was last encountered in the original question-level interaction sequence, before expansion into KC-level interactions. This recency signal is embedded using learnable Fourier features \cite{learn_fourier}. In contrast, positional encoding injects information about the absolute position of each element within a sequence. Unlike positional encoding, recency encoding directly captures temporal patterns relevant to learning, such as forgetting and repetition.


By combining label leakage prevention and recency encoding, our models achieve superior performance compared to their unmodified counterparts. In particular, AKT augmented with our methods consistently outperforms existing baselines. These contributions underscore the importance of both preventing leakage and incorporating recency information for effective knowledge tracing.

To ensure fairness in benchmarking, we standardize the sequence lengths of questions across different models, preventing performance distortions due to sequence-length variations caused by the expanded KC-sequence approach.

In summary, this paper makes the following key contributions:
\begin{itemize}
	\item We empirically demonstrate and quantify the effect of label leakage in widely used KT models.
	\item We propose a simple and computationally efficient method to prevent label leakage using masked ground-truth labels.
	\item We introduce recency encoding, which explicitly encodes the number of steps since each item’s last occurrence in the sequence.
	\item We perform extensive experiments showing that our proposed methods can improve performance across multiple KT models and benchmark datasets.
\end{itemize}


\label{intro}

\section{Related Work}
\label{sec:related}


Bayesian Knowledge Tracing (BKT)\cite{bkt} employs Hidden Markov Models to represent student mastery as binary states (mastered or not mastered). Due to its strong assumption of independence among knowledge concepts (KCs), it inherently avoid the label leakage issue discussed in our work. The same situation applies to other traditional models such as Item Response Theory (IRT) \cite{irt}. However, this assumption severely limits their expressiveness, often leading to inferior predictive performance compared to more complex approaches\cite{Howdeep,whendeep,DKTforget}.

With the advent of deep learning (DL), KT models have gained the capability to capture intricate dependencies in student learning data. The pioneering model in this domain, Deep Knowledge Tracing (DKT), utilizes recurrent neural networks (RNNs) to model sequences of student interactions \cite{dkt}. Subsequently, several variants of DKT have been proposed to enhance its performance or interpretability \cite{dkt_plus,DKTforget}. However, these models typically rely on KC-level expansions of the interaction sequence, making them vulnerable to label leakage, especially when individual items are associated with multiple correlated KCs.

\citet{pykt} highlighted the problem of label leakage during evaluation and proposed methods that simulate realistic production settings to enable fairer model comparisons. However, this approach introduces significant computational overhead. In contrast, we propose simple yet effective techniques that directly eliminate leakage, removing the need for specialized evaluation procedures. Additionally, we quantify the impact of leakage across multiple models.

Transformer-based models, characterized by self-attention mechanisms \cite{transformer}, have become increasingly popular alternatives to RNN-based KT models. Attention mechanisms naturally facilitate the implementation of specialized masks to prevent information leakage within sequences. Prior work, such as \citet{aaai2nd}, has designed customized masks to prevent leakage across grouped questions in datasets where students receive feedback at a group rather than an individual item level. Inspired by these approaches, we investigate integrating masking strategies within transformer-based KT models to explicitly mitigate label leakage while retaining contextual information from previously encountered KCs. However, our simpler \texttt{MASK} encoding strategy showed better performance.

\citet{akt} introduced the AKT model, incorporating a modified scaled dot-product attention mechanism in which the relative distance between items influences the attention weights. They found that adding positional encoding provided no performance benefit. In contrast, our recency encoding explicitly captures time-step distance information and demonstrates improved results.


\section{Background}
\label{sec:background}

Let $Q$ denote the set of all questions. A student's interaction at time step $t$ is represented by the tuple $\left(q_t, r_t\right)$, where $q_t \in Q$ is the question posed and $r_t \in \{0,1\}$ indicates whether the student's response was incorrect ($0$) or correct ($1$). The primary objective of a Deep Learning-based Knowledge Tracing (DLKT) model is to predict the student's response $r'_t$ to the question $q_t$, given the chronological sequence of prior interactions up to time step $t-1$: $\left(q_1, r_1\right), \ldots, \left(q_{t-1}, r_{t-1}\right)$.

Let $C$ be the set of knowledge concepts (KCs). Each question typically relates to multiple KCs, represented by the mapping $m: Q \rightarrow 2^C$, where $2^C$ denotes the power set of $C$. 

\begin{figure}[h]
\centering
\includegraphics[width=0.5\textwidth]{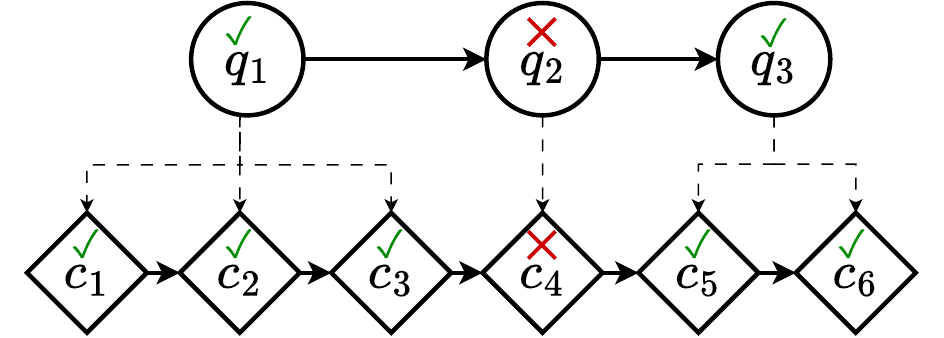}
\caption{Expansion from question--student interaction sequence to KC--student interaction sequences. Green symbols denote correct responses, while red symbols indicate incorrect response. The figure is adapted from \cite{csedu25}}
\label{fig1}
\end{figure}

Several existing models transform the original question-student interaction sequences into KC-student interaction sequences. Specifically, each question interaction is expanded into interactions involving its constituent KCs, as depicted in Figure~\ref{fig1}. Formally, given an ordered set of KCs, consider a question $q$ associated with $n$ KCs. Then, a single interaction $(q, r)$ is expanded to multiple interactions: $(c_1, r), \ldots, (c_n, r)$, where $m(q) = \{c_1, c_2, \ldots, c_n\}$. This procedure increases the sequence length. Note that certain models retain the original question in the expanded sequence, while others remove it entirely.

Since the total number of KCs is generally much lower than the number of questions, this expansion reduces the sparsity issue inherent in student-item interactions \cite{pykt}. Moreover, this strategy can significantly decrease the number of model parameters required \cite{akt}.

In the following subsections, we briefly introduce important DLKT models that use the KC expansion approach. The models we selected do not rely on any features other than KCs and question identifiers. Other models might use additional educational features, such as time spent on an exercise or personal student attributes. However, our goal is to isolate the label leakage caused by KC expansion. Models that incorporate these extra features can have an unfair advantage when comparing different architectural choices.

\subsection{Review of Deep Knowledge Tracing (DKT)}
\label{dkt}
Deep Knowledge Tracing (DKT) introduced the use of deep learning models for the knowledge tracing task \cite{dkt}. The model architecture is based on a recurrent neural network (RNN), with two variants: a vanilla RNN and Long Short-Term Memory (LSTM). In this work, we focus solely on the LSTM-based version. DKT operates on the expanded KC-student interaction sequence and omits question identifiers entirely, predicting future student responses using only the KC-level interactions.

\begin{equation}
    r_t' = DKT\left(c_t ; (c_{t-1}, r_{t-1}), \ldots, (c_1, r_1)\right)
\end{equation}

Each pair $(c, r)$ - a knowledge concept and the student’s binary response - is embedded into a fixed-dimensional vector $e_{(c, r)}$. These embeddings are sequentially passed to the LSTM, which produces a corresponding hidden state $h_t$ at each time step. The hidden states are then fed into a fully connected layer followed by a sigmoid activation:

\begin{equation}
\mathbf{y}_t = \sigma\left(\mathbf{W} h_t + \mathbf{b}\right)
\end{equation}

Here, $\mathbf{W}$ and $\mathbf{b}$ are the weight matrix and bias vector, and $\sigma$ is the sigmoid function. The output $\mathbf{y}_t$ is a vector whose dimensionality matches the number of KCs, with each component representing the predicted probability of a correct response for the corresponding KC. The model’s final prediction is given by $r'_t = \mathbf{y}_t[c_t]$.

In the original formulation, two methods were used to construct the input embeddings $e_{(c, r)}$. The first method applies one-hot encoding to each $(c, r)$ pair, resulting in a vector of size $2|C|$. To avoid the high dimensionality associated with large KC sets, the second method samples a random vector from a standard normal distribution, $e_{(c, r)} \sim \mathcal{N}(\mathbf{0}, \mathbf{I})$, for each pair. Notably, both embedding types are fixed and not learned during training.

More recent DKT implementations \cite{pykt,edustudio} instead learn a unique embedding for each $(c, r)$ pair as trainable parameters. This is functionally similar to using a one-hot encoding followed by a linear projection without bias, producing embeddings of dimension $d$, but at significantly lower computational cost. Following this trend, we also adopt the trainable embeddings approach in our implementation.

\subsubsection{DKT+}
\label{dkt_plus}
DKT+ extends the original DKT by adding regularization terms to address inconsistent predictions and unrealistic fluctuation over time \cite{dkt_plus}. It introduces a reconstruction loss to align predictions with recent observed responses and two smoothness penalties to reduce abrupt changes in predicted knowledge states.



\subsection{Review of Self-Attentive Knowledge Tracing (SAKT)}

Self-Attentive Knowledge Tracing (SAKT) is a transformer-based architecture introduced to overcome limitations of RNN-based knowledge tracing models, particularly their inefficiency in handling sparse interactions and long-range dependencies \cite{sakt}. Unlike RNNs, which inherently capture sequence order through recurrence, attention-based models like SAKT must explicitly encode positional information to maintain the temporal context of student interactions.This is achieved via positional encoding, which adds position-specific information to the input embeddings so that the model can take the order of interactions into account.

In SAKT, each student interaction \( x_t = (c_t, r_t) \) consisting of a KC \( c_t \) and a response \( r_t \). $x_t$ is mapped into a vector representation using a learned interaction embedding matrix \( M \in \mathbb{R}^{2C \times d} \), where \( C \) here is the number of unique KCs and \( d \) is the latent dimension. Note that in SAKT, \( c_t \) can refer to either an exercise or a KC; however, in our paper, we define \( c_t \) to refer strictly to a KC. 

The unanswered question is represented using a separate KC embedding matrix \( E \in \mathbb{R}^{C \times d} \). 

To incorporate temporal ordering, a learned positional embedding matrix \( P \in \mathbb{R}^{n \times d} \) is added element-wise to each interaction embedding:

\begin{equation}
\label{eq:sakt}
\hat{M}_t = M_{x_t} + P_t,
\end{equation}

where \( n \) is the maximum sequence length and \( P_t \) is the positional embedding for timestep \( t \). While $M_{x_t}$ is the encoded $x_t$ using the interaction embedding matrix.



\subsection{Review of Attentive Knowledge Tracing (AKT)}
\label{sec:akt}
\label{akt} 

\begin{figure}[ht]
\centering
\includegraphics[width=0.4\textwidth]{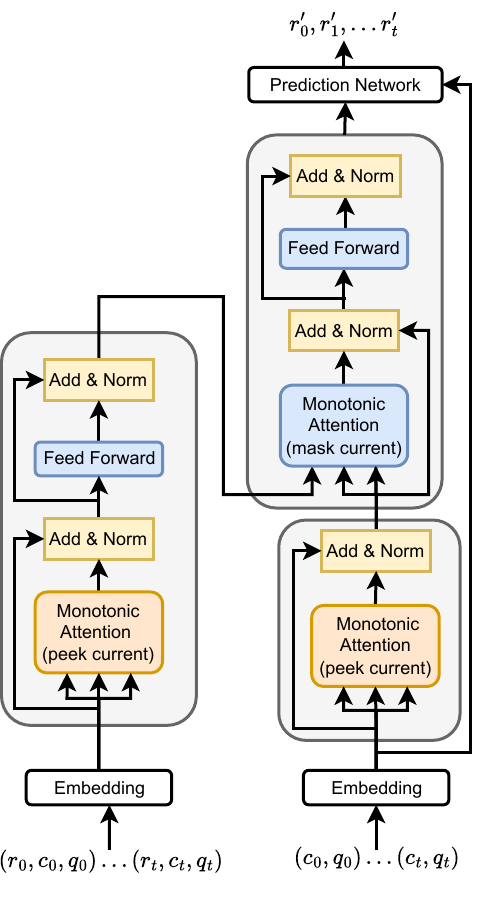}
\caption{Simplified schematic of the AKT model architecture. Note that several attention blocks are repeated in the complete model but omitted here for clarity. In each attention block, inputs are processed from left to right in the order: query, key, and value. Adapted from \cite{csedu25}.
}
\label{fig:akt}
\end{figure}

Unlike standard scaled dot-product attention, \textit{Attentive Knowledge Tracing} (AKT) \cite{akt} employs a modified attention mechanism that incorporates the relative distances between items in the sequence. Attention weights decrease as the distance between items increases, modeling a natural decay. This decay in attention reflects the forgetting behavior observed in human learning. They call this approach monotonic attention.

AKT consists of two distinct self-attention encoders. The first, the question encoder, generates a contextual representation of the current question based on past questions and associated KCs, but without response information. It processes the input sequence $(q_{t-1}, c_{t-1}), \ldots, (q_1, c_1)$ and produces the representation $x_t$ as follows:

\begin{equation}
x_t = f_{qenc}(e_{(q_t, c_t)}, e_{(q_{t-1}, c_{t-1})}, \ldots, e_{(q_1, c_1)})
\end{equation}

where $e_{(q_t, c_t)}$ denotes the embedding of the question-KC pair at time $t$.

The second encoder, the knowledge encoder, captures a contextualized view of the student's knowledge state. It takes as input a sequence of embeddings derived from past responses, KCs, and questions:

\begin{equation}
y_t = f_{kenc}(e_{(r_{t-1}, c_{t-1}, q_{t-1})}, \ldots, e_{(r_1, c_1, q_1)})
\end{equation}

with $e_{(r_t, c_t, q_t)}$ being the embedding of the response-KC-question triplet.

The outputs of both encoders are fed into a knowledge retriever module, which uses another monotonic attention mechanism to retrieve relevant contextual information for answering the current question:

\begin{equation}
h_t = f_{kr}(x_1, \ldots, x_t, y_1, \ldots, y_{t-1})
\end{equation}

The vector $h_t$ is then passed through a feed-forward network to predict the student’s response to a specific KC. A schematic of the AKT architecture is provided in Figure~\ref{fig:akt}.

AKT incorporates two types of attention masks. The first is a lower triangular mask applied to both encoders to prevent any influence from future information, i.e., elements $\left(r_{t+1}, c_{t+1}, q_{t+1}\right), \left(r_{t+2}, c_{t+2}, q_{t+2}\right), \ldots$. The second is a strictly lower triangular mask (zeroing out the diagonal) used in the knowledge retriever to block access to both current and future inputs during prediction. Despite these safeguards, AKT is still susceptible to label leakage, due to its reliance on the KC-response sequence as input—a point further discussed in Section~\ref{sec:leak_problems}.

The model’s embedding strategy is inspired by the Rasch model from item response theory (IRT) \cite{irt}, which estimates a student’s probability of correctly answering a question based on question difficulty and student ability. AKT adapts this idea to construct embeddings for both $(q_t, c_t)$ and $(r_t, c_t, q_t)$ using:

\begin{align}
 e_{(q_t,c_t)} & = e_{c_t} + \mu_{q_t} \cdot d_{c_t} \label{eq:akt_emb1}\\
 e_{(r_t, c_t, q_t)} & = e_{(c_t, r_t)} + \mu_{q_t} \cdot f_{(c_t, r_t)}  \label{eq:akt_emb2}
\end{align}

Here, $d_{c_t}$ and $f_{(c_t, r_t)}$ are "variation vectors" capturing question-KC interactions, and $\mu_{q_t}$ is a scalar representing the difficulty of question $q_t$. The KC embedding is $e_{c_t}$, and the concept-response pair embedding is defined as:

\begin{equation}
    \label{akt:emb}
    e_{(c_t, r_t)} = e_{c_t} + g_{r_t}
\end{equation}

where $g_1$ and $g_0$ correspond to the embeddings for correct and incorrect responses, respectively.

\section{Label Leakage Problems}
\label{sec:leak_problems}

Label leakage problems in Knowledge Tracing models manifest primarily during two critical phases: evaluation and training.

\subsection{Evaluation Problems}
\label{sec:eval}
\begin{figure}[ht]
	\centering
	\includegraphics[width=0.5\textwidth]{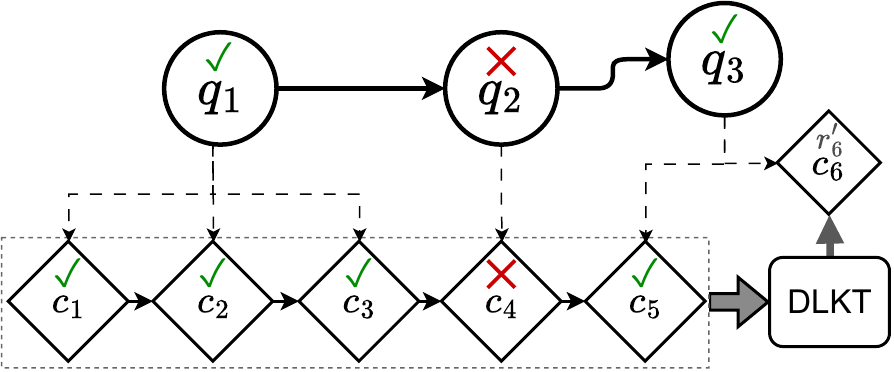}
	\caption{\textit{One-by-one} evaluation method on expanded sequences. Note the leakage risk, as the ground-truth label for $c_5$ could influence prediction $r_6'$ since both belong to question $q_3$. The figure is adapted from \cite{csedu25}}
	\label{fig:onebyone}
\end{figure}

In \cite{pykt}, two methods were proposed for evaluating models that operate on expanded KC-student interaction sequences. The first method, termed "one-by-one" evaluation (see Figure~\ref{fig:onebyone}), treats the expanded KC-level sequence as a standard interaction sequence. Each KC interaction is therefore evaluated individually according to its position in the expanded sequence, without considering which KCs originated from the same question or preserving the original question-level grouping.

\begin{figure}[ht]
	\centering
	\includegraphics[width=0.5\textwidth]{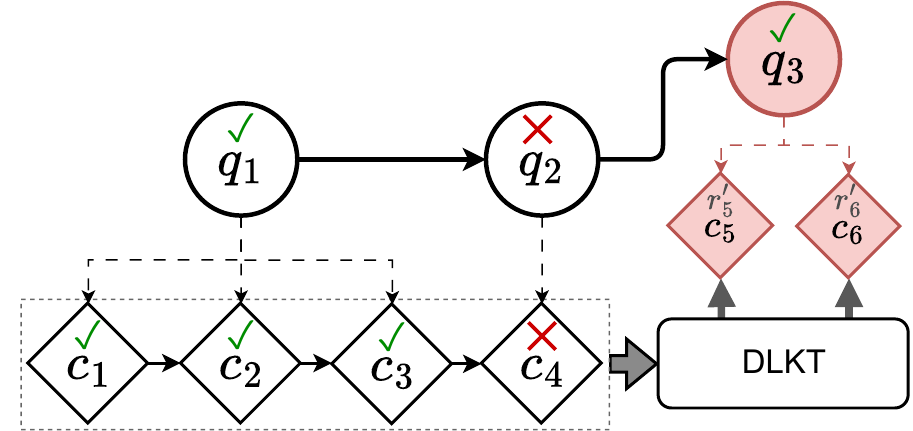}
	\caption{\textit{All-in-one} evaluation method. Predictions for $c_5$ and $c_6$ are independently generated despite belonging to the same question, $q_3$. The figure is adapted from \cite{csedu25}}
	\label{fig:allinone}
\end{figure}

The second method, named "all-in-one" evaluation (Figure~\ref{fig:allinone}), preserves the correspondence between KCs and their original questions during evaluation. For a question associated with multiple KCs, the model produces a separate prediction for each KC independently; that is, the prediction for one KC is not conditioned on the other KCs associated with the same question. The results for these KCs are subsequently aggregated using an aggregation function, commonly the mean, to yield the final prediction per question. In this study, the mean is consistently used for aggregation.

The "one-by-one" evaluation method does not align with real-world production settings because ground-truth labels are not available for individual KCs of unanswered questions. This misalignment leads to label leakage and produces misleading evaluation outcomes \cite{pykt}. Hence, the "all-in-one" evaluation approach is used in this work for models susceptible to ground-truth label leakage.

Nevertheless, the "all-in-one" method requires the model to be evaluated separately for each KC associated with a question to ensure that the KCs are predicted independently. Consequently, the computational cost increases with the number of KCs per question. As a result, this method is impractical for frequent use on the validation set and is typically reserved solely for final testing. This discrepancy between validation and test procedures can mislead model selection strategies such as cross-validation.

Another evaluation challenge arises from the increased sequence length when expanding original sequences into KC-student interactions. Fair benchmark comparisons require consistent evaluation lengths measured in terms of questions rather than KCs. Typically, benchmarks employ the same sequence window for both expanded and original sequences, resulting in unfair comparisons since expanded sequences inherently contain fewer questions. This paper strictly enforces consistent sequence lengths (in terms of questions) across models for equitable benchmarking.

\subsection{Training Problems}
\label{sec:train_problems}

Although the "all-in-one" evaluation method reliably mirrors production environments, label leakage during training remains problematic. In expanded KC-level sequences, the ground-truth response of a question is repeated across all KCs associated with that question. Consequently, when predicting a later KC, the model may learn to rely on the ground-truth response observed for an earlier KC from the same question, rather than learning solely from the student's prior interaction history. Since this response is not available when predicting an unanswered question in production, the learned dependency does not reflect the information available at inference time and may impair generalization. This problem becomes particularly severe for datasets containing multiple KCs per question, as will be further illustrated in Section~\ref{sec:experiments}.

The root cause of label leakage during training is the discrepancy between expanded sequence modeling and the actual data distributions encountered during inference. Consider two interaction events, \((x_i, r_i)\) and \((x_j, r_j)\), within an expanded sequence where \(j > i\). Let \(\mathcal{Q}_{i,j}\) denote the event that both positions \(i\) and \(j\) correspond to the same question, and let \(\mathcal{H}_j\) denote the interaction history up to (but not including) position \(j\). The probability that the responses \(r_i\) and \(r_j\) match can be decomposed as:

\[
	\begin{aligned}
		P(r_j = r_i \mid \mathcal{H}_j) & = P(r_j = r_i \mid \mathcal{H}_j, \mathcal{Q}_{i,j}) \cdot P(\mathcal{Q}_{i,j} \mid \mathcal{H}_j)               \\
		                                & \quad + P(r_j = r_i \mid \mathcal{H}_j, \neg\mathcal{Q}_{i,j}) \cdot P(\neg\mathcal{Q}_{i,j} \mid \mathcal{H}_j)
	\end{aligned}
\]

Since \(P(r_j = r_i \mid \mathcal{H}_j, \mathcal{Q}_{i,j}) = 1\), it follows that:
\[
	P(r_j = r_i \mid \mathcal{H}_j) \geq P(\mathcal{Q}_{i,j} \mid \mathcal{H}_j)
\]

Models trained on expanded interaction sequences can thus implicitly learn the probability \(P(\mathcal{Q}_{i,j} \mid \mathcal{H}_j)\). Consequently, rather than relying on historical interaction patterns, the model may infer the output directly from the previous label $r_i$. At inference time, however, the ground-truth responses for multiple KCs associated with the same unanswered question ($\mathcal{Q}_{i,j}$) are not available, creating a discrepancy between training and inference that can lead to performance degradation.

To address this, we restrict the model to accessing only information that would be available at production time by masking any signals that could lead to label leakage. Specifically, the interaction history used to predict \(r_j\) should be restricted to:

\[
	\mathcal{H}_j \subset \{(x_i, r_i) \mid i < j, \neg\mathcal{Q}_{i,j}\} \cup \{x_i \mid i < j, \mathcal{Q}_{i,j}\}
\]

That is, for earlier interactions involving the same question, only the input \(x_i\) is available, not the response \(r_i\). Thereby, preventing the ground-truth response of the current question from leaking into predictions for other KCs associated with that question.


\section{Preventing Label Leakage}

In this section, we address the problem of label leakage in KT models. We investigated several approaches to mitigate this issue and present the Mask Label method as our primary solution, demonstrating its superiority through extensive experimentation.

\subsection{Mask Label Method (Proposed Approach)}
\label{sec:mask}

\begin{figure}[ht]
\centering
\includegraphics[width=0.5\textwidth]{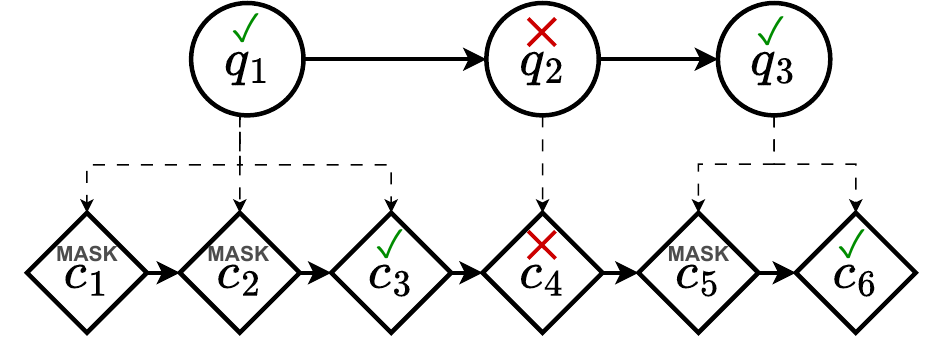}
\caption{Expansion of question--student interaction sequence into KC--student interaction sequence with \texttt{MASK} labels. Green and red symbols represent correct and incorrect responses, respectively. The figure is adapted from \cite{csedu25}}
\label{fig:mask}
\end{figure}

We propose a simple yet effective method to prevent label leakage by incorporating a special mask label, denoted as \texttt{MASK}, into the expanded KC interaction sequence alongside the traditional binary response labels (correct \texttt{1}, incorrect \texttt{0}). Any ground-truth response preceding another KC within the same question is replaced with \texttt{MASK}, explicitly preventing intra-question label leakage. Figure \ref{fig:mask} illustrates this method clearly.

For instance, a question $q$ associated with KCs $(c_1, c_2, c_3)$ and a response $r$ would be represented in an expanded sequence as:
\[
\ldots, \left(q, c_1, \texttt{MASK}\right), \left(q, c_2, \texttt{MASK}\right), \left(q, c_3, r\right), \ldots
\]
In this approach, only the final KC retains its actual ground-truth label.

The \texttt{MASK} token is exclusively applied to the model’s input embeddings, whereas the output predictions are constrained to the original labels \texttt{0} and \texttt{1}. This procedure is applied during both training and inference, eliminating the need for special-case handling at inference time. 

The \texttt{MASK} labels are assigned only once during construction of the expanded sequence; the sequence is not re-masked separately for each prediction. Incorporating masked labels into embeddings therefore requires only minimal changes, making this approach broadly applicable across different architectures.

We refer to models that adopt this method with the suffix “-ML” (mask label), and in the following subsections, we present a set of model variants that implement this approach, namely DKT-ML, DKT-ML+, AKT-ML, and SAKT-ML. 

\subsubsection{Separate KC and response Embeddings (AKT-ML and DKT-ML)}
\label{sec:rasch-ml}

AKT employs separate embeddings for correct ($g_1$) and incorrect ($g_0$) responses. To incorporate the mask label, we introduce an additional embedding vector, $g_{\texttt{MASK}}$, which is learned jointly with the existing response embeddings. The resulting embedding for a KC $c_t$ with a masked response label is
\[
e_{(c_t, \texttt{MASK})} = e_{c_t} + g_{\texttt{MASK}}.
\]

We apply the same strategy to DKT by using separate embeddings for correct ($g_1$), incorrect ($g_0$), and masked ($g_{\texttt{MASK}}$) response labels. We refer to this variant as \textbf{DKT-ML} (DKT with Masked Labels).

AKT requires an additional "variation" embedding, we therefore introduce a distinct variation embedding, $f_{(c_t, \texttt{MASK})}$, for each KC paired with a masked response label. We refer to this variant as \textbf{AKT-ML} (AKT with Masked Labels).

\subsubsection{Combined KC and response embedding  (DKT-ML+ and SAKT-ML)}

For DKT+, and SAKT, we encode KC-response pairs into single embeddings using a unified input embedding matrix. To incorporate the \texttt{MASK} label, we create a new embedding vector for each KC that represent a KC with a \texttt{MASK} label. Thus, each (KC, response) pair—including masked entries—is directly mapped to its corresponding embedding vector from this larger embedding matrix.

These embeddings are then fed into the respective model architectures (RNN for DKT+ and attention for SAKT). Based on this design, we define DKT-ML+ and SAKT-ML as variants of their original counterparts.

Note that this encoding choice increases the number of parameters, as each input requires a separate embedding vector for the \texttt{MASK} label.

\subsection{Alternative Methods Investigated}

In addition to the Mask Label method, we have rigorously explored several alternative strategies to address label leakage. Each method is described below, including their technical implementation and potential limitations.

\subsubsection{Averaged Embeddings (DKT-Fuse, AKT-Fuse)}
\label{sec:fuse}

This method avoids KC-student expansion by aggregating embeddings of KCs within the same question into a single representative embedding. This has been used in \cite{qikt,iekt} to incorporate KCs alongside question embeddings. Formally, the embedding for a given question $q$ with response $r$ is computed as:
\[
\bar{e}_{(q, r)} = \frac{1}{|m(q)|}\sum_{c \in m(q)} e_{(c,r)}
\]
Averaging the KC embedding vectors makes it difficult to incorporate additional information about individual KCs, such as their recency. 

We create two model variants to test this approach, DKT-Fuse and AKT-Fuse. Our experiments demonstrate that it underperforms compared to our Mask Label method and AKT-Fuse was the worst performing model variant on a single benchmark, as seen in Section~\ref{sec:leak_lab}.

\subsubsection{Autoregressive Decoding (DKT-AD)}
\label{sec:ad}

The Autoregressive Decoding method, DKT-AD, addresses label leakage by sequentially generating responses for KCs within the same question based on model predictions rather than ground-truth labels. Specifically, given a question with KCs $(c_1, c_2, c_3)$ and true response $r$, the expanded sequence is modeled as:
\[
\ldots,\left(q, c_1, r'_1\right), \left(q, c_2, r'_2\right),\left(q, c_3, r\right),\ldots
\]
where $r'_1$ and $r'_2$ are model-generated predictions. This approach ensures that the model's inputs reflect realistic inference conditions by preventing direct access to ground-truth labels during training. Despite its theoretical appeal, autoregressive decoding introduces potential compounding errors due to reliance on model predictions. In our experiments, DKT-AD exhibited diminished accuracy compared to the Mask Label approach. Moreover, it is only applicable to RNN architectures.

\subsubsection{Special Attention Masking (AKT-QM)}

This method modifies the attention mechanism of attention-based KT models, such as AKT, to explicitly prevent interactions between KCs belonging to the same question. We apply this modification to AKT and refer to the resulting variant as AKT with Question-level Masking (AKT-QM). 

The modified attention mask is defined as:
\[
A_{ij} = 
\begin{cases}
0 & \text{if } \text{KCs } i \text{ and } j \text{ belong to the same question}, \\
0 & \text{if } i \leq j, \\
1 & \text{otherwise}
\end{cases}
\]

This formulation effectively prevents information flow within intra-question KCs, thus mitigating leakage. However, this masking scheme is computationally expensive and it is only applicable to attention-based models. Moreover, empirical results consistently shows inferior performance relative to the Mask Label method.


\section{Recency Encoding for Knowledge Tracing Models}

Transformer-based architectures commonly incorporate positional encodings to capture the order of elements in a sequence. However, the AKT model~\cite{akt} omits such encoding, instead relying on a monotonic attention mechanism that leverages relative distances between all items in the sequence. Empirically, AKT outperforms variants that use positional encoding, as shown in \cite{akt}. The authors assume that, unlike natural language tasks where long-range dependencies are frequent, student learning behavior is more influenced by recent interactions\cite{akt}.

Motivated by this insight, we propose a method that explicitly encodes the number of time steps since the most recent occurrence of the same KC within a student's interaction sequence. This distance provides a meaningful pedagogical signal, which is important for modeling the effects of forgetting and repetition. We refer to this method as \emph{recency encoding}.

Rather than using hand-crafted or fixed encoding, we adopt an approach based on \emph{learnable Fourier features}~\cite{learn_fourier}. Given a scalar distance \( d \in \mathbb{R} \), the encoding is defined as:

\begin{equation}
	\gamma(d) = \left[ \cos(d \mathbf{w}_f + \mathbf{b}_f),\ \sin(d \mathbf{w}_f + \mathbf{b}_f) \right],
\end{equation}

where \( \mathbf{w}_f \in \mathbb{R}^{\frac{D}{2}} \) is a trainable frequency vector and \( \mathbf{b}_f \in \mathbb{R}^{\frac{D}{2}} \) is a learnable bias. Both are initialized from normal distributions with unit variance: \( \mathbf{w}_f \sim \mathcal{N}(0, 1) \) and \( \mathbf{b}_f \sim \mathcal{N}(0, 1) \). For the first occurrence, we set \( d = 0 \).

The resulting Fourier features are passed through a feedforward transformation:
\begin{equation}
	\mathrm{DE}(d) = \phi(\gamma(d)) W_p + b_p,
\end{equation}
where $W_p \in \mathbb{R}^{D \times D'}$ is a learnable projection matrix, $b_p \in \mathbb{R}^{D'}$ is a learnable bias vector, and $D'$ is the dimension of the input embedding. The transformation $\phi$ is implemented as a linear layer followed by a Gaussian Error Linear Units (GeLU) activation function, following the design in~\cite{learn_fourier}.

This architecture allows the model to generalize to unseen or infrequent distances, which may be sparsely represented in the training data. We incorporate the proposed recency encoding into three existing KT models, denoting each variant with a superscript $d$:



\begin{itemize}
	\item \textbf{SAKT-ML\textsuperscript{d}}: Replaces the original positional encoding in Equation~\ref{eq:sakt} with the proposed recency encoding.
	\item \textbf{DKT-ML\textsuperscript{d}}: Adds recency encoding to the input embeddings before they are passed to the recurrent layer. Specifically, the recency encoding is added to $e_{(c_t, \texttt{0})}$, $e_{(c_t, \texttt{1})}$, and $e_{(c_t, \texttt{MASK})}$.
	\item \textbf{AKT-ML\textsuperscript{d}}: Adds recency encoding only to the value vectors in the attention mechanism, while leaving the query and key vectors unchanged. The original monotonic attention mechanism of AKT is preserved, as described in Section~\ref{sec:akt}.

\end{itemize}

\section{Experiments}
\label{sec:experiments}
\begin{table}[ht]
	\caption{Dataset attributes after preprocessing. "ques.", "KCs", "studs.", and "KCs/ques." refer to the number of questions, knowledge concepts, students, and average KCs per question, respectively. "KC-grps." indicates the number of unique KC groups per question. This table is adapted from \cite{csedu25}}
	\label{tab:datasets}
	\begin{tabular}{@{}llllll@{}}
		\toprule
		\textbf{Dataset} & \textbf{ques.} & \textbf{KCs} & \textbf{studs.} & \textbf{KC-grps.} & \textbf{KCs/ques.} \\
		\midrule
		Algebra2005      & 173650         & 112          & 574             & 263               & 1.353              \\
		ASSISTments2009  & 17751          & 123          & 4163            & 149               & 1.196              \\
		CorrAS09         & 17751          & 246          & 4163            & 149               & 2.393              \\
		Duolingo2018     & 694675         & 2521         & 2638            & 7883              & 2.702              \\
		Riiid2020        & 13522          & 188          & 3822            & 1519              & 2.291              \\
		\bottomrule
	\end{tabular}
\end{table}

Since our models rely on KCs, we selected datasets that exhibit a variety of KC-related characteristics, including KC cardinality and the mean number of KCs associated with each question. We also report the number of distinct KC combinations per question, where each combination defines a unique group of KCs tied to a specific question. Table~\ref{tab:datasets} summarizes these dataset properties. In preprocessing, we retained only the KCs, question identifiers, student identifiers, and the chronological order of interactions, discarding all other features. The datasets employed in our experiments are:

\begin{itemize}
	\item \textbf{ASSISTments2009}\footnote{Available at \url{https://sites.google.com/site/assistmentsdata/home/}}: This dataset originates from the ASSISTments platform and includes student interactions collected between 2009 and 2010. We specifically use the skill-builder version.
	\item \textbf{Riiid2020} \cite{riiid}: Sourced from an AI-based tutoring system, this dataset contains over 100 million interactions. For our experiments, we restrict the data to the first one million interactions. Note that although the dataset was originally intended for a competition and includes additional features, we exclude them to maintain consistent evaluation conditions.
	\item \textbf{Algebra2005} \cite{algebra05}: This dataset was used in the 2010 KDD Cup Educational Data Mining Challenge. We focus only on the subset labeled "Algebra I 2005-2006".
	\item \textbf{Duolingo2018}\cite{duolingo18}: Collected via the Duolingo language-learning application, this dataset includes data from roughly 6,000 learners across multiple languages. We select only those students with English as their primary language who were learning Spanish. In this context, we treat word tokens as KCs, noting that some words may comprise multiple tokens.
\end{itemize}

To further demonstrate the effect of label leakage during training, we construct a synthetic variant of the \textit{ASSISTments2009} dataset, referred to as \textit{CorrAS09}. Both datasets share identical student interactions and question sequences; however, they differ in the set of KCs assigned to each question. In \textit{CorrAS09}, we artificially increase KC correlation by replacing each original KC with a pair of pseudo-duplicates. Specifically, for every KC $c$, we generate two new KCs, $m'(c)$ and $m''(c)$, where $m'$ and $m''$ are mapping functions with disjoint ranges, each matching the size of the original KC set. As a result, every question is annotated with at least two highly correlated (duplicated) KCs, allowing us to amplify the impact of label leakage in a controlled setting.

\subsection{Baseline Models}

To evaluate the effectiveness of our proposed methods, we compare against two groups of baseline models: (1) models that are inherently unaffected by label leakage due to their design choices, and (2) established KT models that utilize KC-level sequence expansions, which may introduce label leakage. The latter group (DKT, DKT+, SAKT, and AKT) has already been introduced in section~\ref{sec:background}

\paragraph{Leakage-Resistant Models.} These models operate without requiring the expansion of questions into individual KC-response pairs. As a result, they are unaffected by label leakage.

\begin{itemize}
    \item \textbf{Dynamic Key-Value Memory Networks (DKVMN)}~\cite{dkvmn}: A memory-augmented neural architecture that stores representations of knowledge concepts in a key-value memory structure.
    \item \textbf{DeepIRT}~\cite{deepirt}: Combines deep learning with item response theory (IRT) to model student ability and question difficulty.
    \item \textbf{Question-centric Interpretable Knowledge Tracing (QIKT)}~\cite{qikt}: A question-centric architecture that incorporates KCs by averaging their embeddings. They use an IRT-based prediction mechanism to generate interpretable outputs.
\end{itemize}


\subsection{Training and Evaluation Setup.} All models were trained using the ADAM optimizer~\cite{adam} with a learning rate of $10^{-3}$. DKT-based models used a batch size of $128$, while all other models were trained with a batch size of $24$. Unless specified otherwise, we performed 5-fold cross-validation on $80\%$ of the students, reserving the remaining $20\%$ for testing. Model performance was evaluated at the question level using the area under the ROC curve (AUC) as the primary metric.

To ensure a fair comparison, we fixed the input sequence length to a maximum of 150 questions for all models. For models that expand question sequences into KC-level sequences, this expansion was applied after restricting the question window size. In these models, each KC generates an individual prediction, and the final prediction for a question is computed as the average of the model's outputs across all associated KCs.

\subsection{Label Leakage Effects and Comparison of Mitigation Strategies}
\label{sec:leak_lab}

To systematically assess the effect of label leakage, we compare models trained with and without mitigation strategies on the ASSISTments2009 and CorrAS09 datasets. As seen in Table~\ref{tab:leakage_table}, the original models (e.g., DKT, AKT) suffer significant performance degradation on CorrAS09 due to duplicated and correlated KCs, revealing their susceptibility to leakage. This contrasts with variants such as DKT-ML, AKT-ML, SAKT-ML, and DKT-ML+, which maintain high and consistent performance across both datasets.

We further benchmarked several alternative mitigation techniques: DKT-Fuse (averaged KC embeddings), DKT-AD (autoregressive decoding), and AKT-QM (special attention masking). While all of these approaches successfully mitigate label leakage, the \texttt{MASK} label method offers several practical advantages. First, it is simpler to implement and integrate into a wide range of model architectures. Second, it is computationally cheaper compared to autoregressive decoding or attention-masking schemes. As shown in Table~\ref{tab:leakage_methods_full}, \texttt{MASK}-based models (e.g., DKT-ML and AKT-ML) outperform other mitigation strategies across most datasets. The performance gain is especially significant for AKT, where the AUC improvements are consistently larger. This combination of ease of use, efficiency, and solid empirical performance makes the \texttt{MASK} label strategy a practical and effective choice for addressing label leakage.



\begin{table}[ht]
\centering
\caption{Effect of label leakage and mitigation methods on model performance. Shown are AUC scores ($\text{mean} \pm \text{std}$) on ASSISTments2009 and CorrAS09 datasets. ML = Mask Label. Some results are adapted from \cite{csedu25}}
\label{tab:leakage_table}
\begin{tabular}{@{}lcc@{}}
\toprule
\textbf{Model} & \textbf{ASSISTments09} & \textbf{CorrAS09} \\
\midrule
DKT           & $0.6990 \pm 0.0007$ & $0.6312 \pm 0.0014$ \\
DKT-ML        & $0.7185 \pm 0.0003$ & $0.7163 \pm 0.0006$ \\
DKT-AD        & $0.7180 \pm 0.0005$ & $0.7148 \pm 0.0002$ \\
DKT-Fuse      & $0.7066 \pm 0.0005$ & $0.7074 \pm 0.0008$ \\
DKT+          & $0.7010 \pm 0.0011$ & $0.6403 \pm 0.0030$ \\
DKT-ML+       & $0.7200 \pm 0.0010$ & $0.7188 \pm 0.0009$ \\
AKT           & $0.7334 \pm 0.0017$ & $0.6361 \pm 0.0020$ \\
AKT-ML        & \textbf{0.7543 $\pm$ 0.0010 }& \textbf{0.7552 $\pm$ 0.0010} \\
AKT-QM        & $0.7193 \pm 0.0134$ & $0.7368 \pm 0.0011$ \\
AKT-Fuse      & $0.7488\pm0.0021$ &  $0.7476\pm0.0011$ \\
SAKT          & $0.6946 \pm 0.0014$ & $0.6336 \pm 0.0062$ \\
SAKT-ML       & $0.7166 \pm 0.0014$ & $0.7189 \pm 0.0009$ \\
QIKT (no leakage) & $0.7472 \pm 0.0008$ & $0.7484 \pm 0.0007$ \\
\bottomrule
\end{tabular}
\end{table}

\begin{table}[ht]
\centering
\caption{Comparison of label leakage mitigation methods across benchmark datasets. Best-performing model in each group is highlighted. The marker * indicates that not all folds were tested (one fold used: train 64\%, validation 16\%, test 20\%).}
\label{tab:leakage_methods_full}
\begin{tabular}{@{}lcccc@{}}
\toprule
\textbf{Model} & \textbf{Algebra05} & \textbf{Riiid20} & \textbf{Duolingo18} \\
\midrule
\textbf{DKT Variants} \\
DKT-ML       & $0.8178 \pm 0.0003$ & $0.6568 \pm 0.0004$ & 0.8681 $\pm$ 0.0004 \\
DKT-AD       & $0.8161 \pm 0.0003$ & $0.6554 \pm 0.0003$ & $0.8679 \pm 0.0003$ \\
DKT-Fuse     & $0.8175 \pm 0.0003$ & $0.6491 \pm 0.0003$ & $0.8786 \pm 0.0001$ \\
\midrule
\textbf{AKT Variants} \\
AKT-ML       & \textbf{0.8282 $\pm$ 0.0011} & \textbf{0.7411 $\pm$ 0.0007} & \textbf{0.8807 $\pm$ 0.0013} \\
AKT-QM       & $0.7919 \pm 0.0072$ & $0.7289 \pm 0.0034$ & $0.8052$* \\
AKT-Fuse     & $0.8220\pm0.0024$ & $0.7393\pm0.0011$ & $0.7682\pm0.0028$ \\
\bottomrule
\end{tabular}
\end{table}

\begin{figure}[h]
\centering
\includegraphics[width=0.95\textwidth]{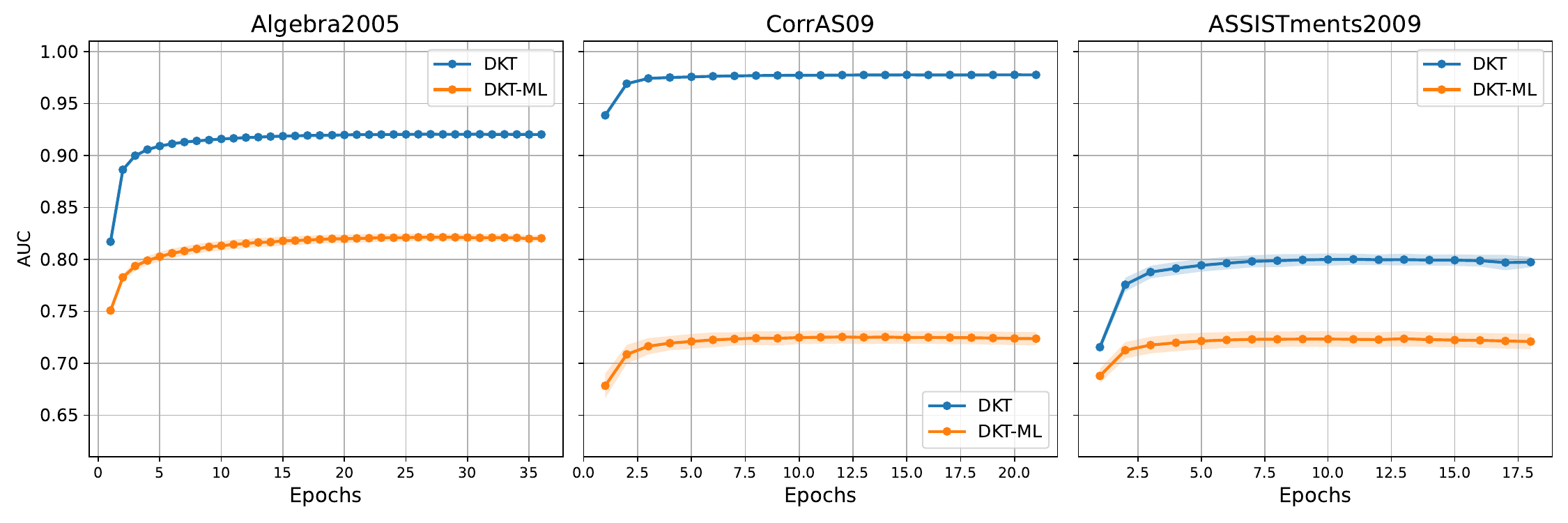}
\caption{Validation loss on the CorrAS09, Algebra2005, and ASSISTments2009 datasets. The \textit{one-by-one} method is used for the original DKT model and DKT-ML, our variant of DKT with added \texttt{MASK} label. DKT shows inflated results due to label leakage, especially in datasets with a higher number of KCs per question (Algebra2005 and CorrAS09). DKT-ML demonstrates similar performance on CorrAS09 and ASSISTments2009. The figure is adapted from \cite{csedu25}}
\label{fig:train}
\end{figure}

Lastly, as we mentioned in Section~\ref{sec:eval}, the expensive but realistic "all-in-one" evaluation method that is designed to mimic production settings diverges significantly from the "one-by-one" method used during model training and selection. This discrepancy introduces misleading validation outcomes due to label leakage. For instance, during training on CorrAS09, we observe that the AUC scores are abnormally higher than those on the original ASSISTments2009 dataset, even though the question data remains identical. This inflation occurs because leakage enables models like DKT to overfit to intra-question KC correlations. Figure~\ref{fig:train} illustrates this phenomenon: DKT achieves unrealistically high validation scores on CorrAS09 and Algebra2005 (datasets with high KC per question), whereas DKT-ML, which applies the \texttt{MASK} label strategy, yields consistent validation performance across both CorrAS09 and ASSISTments2009. This underscores the importance of using leakage-aware strategies not only during evaluation but also throughout the training pipeline.

Among all evaluated techniques, the \texttt{MASK} label method (used in -ML variants) consistently delivers the highest performance across datasets. It outperforms other mitigation strategies such as DKT-AD, DKT-Fuse, and AKT-QM in our experiments, while also being computationally efficient. These results highlight its effectiveness and practicality for addressing label leakage in knowledge tracing models.
\subsection{Effect of Recency Encoding in KT Models}

In this section, we evaluate the impact of augmenting input embeddings with recency encoding, focusing on models that already address label leakage through the mask label (ML) mechanism. Models incorporating recency encoding are denoted with a superscript d (SAKT-ML\textsuperscript{d}, AKT-ML\textsuperscript{d}, and DKT-ML\textsuperscript{d}).

As shown in Table~\ref{tab:cross_combined}, adding recency encoding can provide an advantage on different benchmarks for the three model families (AKT-ML, SAKT-ML, and DKT-ML) . The attention-based model variants achieve the best AUC scores in all benchmarks. These improvements  suggest that temporal recency provides a useful signal. Moreover, recency encoding performed better than positional encoding, as SAKT-ML uses positional encoding while  SAKT-ML\textsuperscript{d} uses only recency encoding.

Unlike attention based models, RNNs do not need positional encoding. However, DKT-ML\textsuperscript{d} gained noticeable improvement on the Duolingo2018 dataset, outperforming the original attention based models without recency-encoding. We hypothesise that this improvement stems from the stronger influence of forgetting in the context of language learning, which makes recency information particularly valuable.

In addition to the learnable encoding, we also evaluated a fixed Fourier-based encoding as described in NeRF~\cite{NeRF}, where input distances are projected using predefined frequency bands. As shown in Table~\ref{tab:fix}, the fixed variant (denoted with the superscript \( d\text{-fix} \)) yields improvements over the base models in several cases. However, these gains are generally smaller and less consistent than those achieved with the learnable encoding.


\begin{table}[ht]
\caption{Cross-Validation Performance of AKT-based and DKT-based Models}\label{tab:cross_combined}%
\begin{tabular}{@{}llcccc@{}}
\toprule
Model Type & Model & ASSISTments09 & Algebra05 & Riiid20 & Duolingo2018 \\
\midrule
\multirow{2}{*}{AKT-based} 
& \textbf{AKT-ML\textsuperscript{d}} & $\mathbf{0.7566\pm0.0006}$ & $\mathbf{0.8325\pm0.0023}$ & $\mathbf{0.7441\pm0.0012}$ & $\mathbf{0.8947\pm0.0017}$ \\
& AKT-ML & $0.7543\pm0.0010$ & $0.8282\pm0.0011$\textsuperscript{\dag} & $0.7411\pm0.0007$ & $0.8807\pm0.0013$ \\
\midrule
\multirow{2}{*}{SAKT-based} &
\textbf{SAKT-ML\textsuperscript{d}}   & $\mathbf{0.7191 \pm 0.0012}$ &  $\mathbf{0.8082 \pm 0.0053}$ & $\mathbf{0.6522 \pm 0.0003}$ & $\mathbf{0.8687 \pm 0.0008}$ \\
& SAKT-ML    & $0.7166 \pm 0.0014$ &  $0.7954 \pm 0.0005$ & $0.6460 \pm 0.0007$ & $0.8472 \pm 0.0014$ \\
\midrule
\multirow{2}{*}{DKT-based} 
& \textbf{DKT-ML\textsuperscript{d}} & $\mathbf{0.7202\pm0.0018}$ & $\mathbf{0.8179\pm0.0009}$ & $0.6560\pm0.0003$ & $\mathbf{0.8901\pm0.0008}$ \\
& DKT-ML & $0.7185\pm0.0003$ & $0.8178\pm0.0003$ & $\mathbf{0.6568\pm0.0004}$ & $0.8681\pm0.0004$ \\
\botrule
\end{tabular}
\footnotetext{Source: Cross-validation AUC scores of AKT-based and DKT-based models on various datasets. Best results per group are highlighted.}
\end{table}

\begin{table}[ht]
\caption{Performance Comparison of AKT-based and DKT-based Models}\label{tab:fix}%
\begin{tabular}{@{}llcccc@{}}
\toprule
Model Type & Model & ASSISTments09 & Algebra05 & Riiid20 & Duolingo2018 \\
\midrule
\multirow{3}{*}{AKT-based} 
& \textbf{AKT-ML\textsuperscript{d-fix}} & $0.7525$ & $0.8181$ & $0.7431$ & $0.8930$ \\
& \textbf{AKT-ML\textsuperscript{d}}     & $\mathbf{0.7549}$ & $\mathbf{0.8346}$ & $\mathbf{0.7443}$ & $\mathbf{0.8943}$ \\
& \textbf{AKT-ML}                        & $0.7530$ & $0.8285$ & $0.7422$ & $0.8839$ \\
\midrule
\multirow{3}{*}{DKT-based} 
& \textbf{DKT-ML\textsuperscript{d-fix}} & $0.7193$ & $0.8181$ & $\mathbf{0.6579}$ & $0.8788$ \\
& \textbf{DKT-ML\textsuperscript{d}}     & $\mathbf{0.7195}$ & $0.8179$ & $0.6556$ & $\mathbf{0.8890}$ \\
& \textbf{DKT-ML}                        & $0.7181$ & $\mathbf{0.8194}$ & $0.6550$ & $0.8683$ \\
\botrule
\end{tabular}
\footnotetext{Source: Performance scores (AUC) of AKT-based and DKT-based models on various datasets. Best results per group are highlighted.}
\end{table}

\subsection{Performance Comparison with Baseline Models}
 
\begin{table}[ht]
\caption{AUC performance results across different model types. 
The marker \dag\ denotes models performing very close to the best-performing model (near tie). The marker ** indicates an impractical test due to the large number of questions in models where question embeddings are used.}
\label{tab:fullbench}
\begin{tabular}{@{}lllll@{}}
\toprule
\textbf{Model} & \textbf{ASSIST09}  & \textbf{Algebra05} & \textbf{Riiid20} & \textbf{Duolingo2018} \\
\midrule
DKT & $0.6990\pm0.0007$ & $0.8070\pm0.0004$ & $0.5961\pm0.0003$ & $0.6518\pm0.0013$ \\
DKT-ML & $0.7185\pm0.0003$ & $0.8178\pm0.0003$ & $0.6568\pm0.0004$ & $0.8681\pm0.0004$ \\
\textbf{DKT-ML\textsuperscript{d}} & $0.7202\pm0.0018$ & $0.8179\pm0.0009$ & $0.6560\pm0.0003$ & $\mathit{0.8901\pm0.0008}$ \\
DKT-ML+    & $0.7200 \pm 0.0010$ & $0.8125 \pm 0.0006$ & $0.6469 \pm 0.0006$ & $0.8683 \pm 0.0008$ \\
SAKT-ML    & $0.7166 \pm 0.0014$ & $0.7954 \pm 0.0005$ & $0.6460 \pm 0.0007$ & $0.8472 \pm 0.0014$ \\
SAKT-ML\textsuperscript{d}   & $0.7191 \pm 0.0012$ & $0.8082 \pm 0.0053$ & $0.6522 \pm 0.0003$ & $0.8687 \pm 0.0008$ \\
AKT & $0.7334\pm0.0017$ & $0.7591\pm0.0045$ & $0.6136\pm0.0015$ & $0.7017\pm0.0183$ \\
AKT-ML & $\mathit{0.7543\pm0.0010}$ & $\mathit{0.8282\pm0.0011}$\textsuperscript{\dag} & $\mathit{0.7411\pm0.0007}$ & $0.8807\pm0.0013$ \\
\textbf{AKT-ML\textsuperscript{d}} & \bm{$0.7566\pm0.0006$} &  \bm{$0.8325\pm0.0023$} & \bm{$0.7441\pm0.0012$} & \bm{$0.8947\pm0.0017$} \\

QIKT & $0.7472\pm0.0008$ & $\mathit{0.8290\pm0.0007}$ & $0.7306\pm0.0005$ & \multicolumn{1}{c}{$-$\textsuperscript{**}} \\
DeepIRT & $0.7215\pm0.0010$ & $0.7779\pm0.0003$ & $0.7312\pm0.0002$ & \multicolumn{1}{c}{$-$\textsuperscript{**}} \\
DKVMN & $0.7215\pm0.0011$ & $0.7768\pm0.0003$ & $0.7327\pm0.0003$ & \multicolumn{1}{c}{$-$\textsuperscript{**}}  \\
\bottomrule
\end{tabular}
\footnotetext{Some results are adapted from \cite{csedu25}}
\end{table}


Table~\ref{tab:fullbench} summarizes the AUC performance across four benchmark datasets. Our proposed model variants consistently outperform their original counterparts on all benchmarks. Notably, \textbf{AKT-ML\textsuperscript{d}} achieves the best overall performance. This is particularly advantageous, as our model variants use significantly fewer parameters compared to question-centric approaches such as QIKT, DeepIRT, and DKVMN, which allocate more parameters per question.

 In most cases, \textbf{AKT-ML} ranks second, demonstrating that the mask label strategy alone provides substantial improvements. Notably, on the \textbf{Duolingo2018} dataset, which exhibits a high number of KCs per question, \textbf{DKT-ML\textsuperscript{d}} is the second-best performing model.

Improvements are particularly pronounced on datasets with a higher average number of KCs per question:  \textbf{Riiid2020} and \textbf{Duolingo2018}. This highlights the effectiveness of our label leakage mitigation. Even on datasets with fewer KCs per question, namely \textbf{ASSISTments2009} and \textbf{Algebra2005}, our variants remain competitive and often lead.



Overall, these results collectively demonstrate that mitigating label leakage and encoding recency information yield significant and robust gains across a wide range of KT benchmarks.

\subsection{Scope and Limitations}

First, our study focuses on KT settings with explicit KC annotations, and the effectiveness of the proposed methods may therefore depend on the quality and granularity of these annotations. Second, our preprocessing retains only question identifiers, KCs, student identifiers, responses, and interaction order. While this controlled setup facilitates the evaluation of label leakage and isolates the effects of the proposed methods, the reported performance may differ when additional contextual features are incorporated. Finally, although our recency encoding can represent discretised elapsed time, we deliberately define recency in our experiments as the number of interactions since a KC was last observed. This choice is consistent with our controlled experimental setting and avoids introducing additional contextual information, such as absolute timestamps.

\section{Conclusion}

In this study, we identified and addressed a significant issue of label leakage in knowledge tracing (KT) models. We proposed a simple yet effective masking strategy that mitigates this leakage during both training and inference. Additionally, we introduced a novel recency encoding mechanism that captures the temporal distance between KC encounters, enabling models to better account for temporal learning dynamics such as forgetting.

Both contributions operate solely at the embedding level, making them computationally efficient and broadly applicable across various KT architectures. Our experimental results consistently demonstrate that these methods outperform their original counterparts, particularly on datasets with multiple KCs per question. 

Notably, the combination of masking and recency encoding yielded the best results across all benchmarks when applied to AKT. Furthermore, our recency encoding outperforms traditional positional encoding in attention-based models like SAKT, reinforcing its utility in modeling learning dynamics.

Overall, our findings show that addressing label leakage and incorporating recency information can substantially improve KT performance with only minimal architectural changes. This makes the proposed methods practical and broadly applicable to a wide range of KT models.

\subsection*{Acknowledgments}
We would like to thank the anonymous reviewers for their helpful feedback and suggestions.

\section*{Declarations}

\subsection*{Competing Interests}
On behalf of all authors, the corresponding author states that there is no conflict of interest.

\subsection*{Funding Information}
Funded by the federal state of Baden-Württemberg (grant number BW6 10).

\subsection*{Author Contribution}
\textbf{Yahya Badran:} Conceptualization, Methodology, Investigation, Data Curation, Formal Analysis, Software, Validation, Visualization, Writing -- Original Draft.\\
\textbf{Christine Preisach:} Writing -- Review \& Editing, Supervision, Funding Acquisition, Conceptualization, Project Administration.

\subsection*{Data Availability Statement}
All data used are publicly available open-access datasets.

\subsection*{Research Involving Human and/or Animals}
Not applicable.

\subsection*{Informed Consent}
Not applicable.


\bibliography{sn-bibliography}

\end{document}